\documentclass[a4paper,fleqn,usenatbib]{mnras}

\usepackage{newtxtext,newtxmath}
\usepackage[T1]{fontenc}
\usepackage{ae,aecompl}

\usepackage{graphicx}	% Including figure files
\usepackage{amsmath}	% Advanced maths commands
\usepackage{mathtools}
\usepackage{hyperref}

\title[Average annual total sunspot area in the last 410 years]{Average annual total sunspot area in the last 410 years: The most probable values and limits of their uncertainties}

\author[Yu. A. Nagovitsyn \& A. A. Osipova]{
Yury A. Nagovitsyn$^{1,2}$\thanks{E-mail: nag@gaoran.ru},
Aleksandra A. Osipova$^{1}$
\\
$^{1}$Central Astronomical Observatory of the Russian Academy of Sciences at Pulkovo, Pulkovskoe sh. 65/1, St Petersburg 196140, Russia\\
$^{2}$St Petersburg State University of Aerospace Instrumentation, Bol'shaya Morskaya ul. 67, St Petersburg 190000, Russia\\
}

\date{Accepted XXX. Received YYY; in original form ZZZ}

\pubyear{2021}

\begin{document}
\label{firstpage}
\pagerange{\pageref{firstpage}--\pageref{lastpage}}
\maketitle

\begin{abstract}
The aim of this work is to create a long (410 years) series of average annual total sunspot areas $AR$ – physically-based index of sunspot activity. We use telescopic observations of the $AR$ index in 1832-1868 and 1875-2020, as well as the relationship between $AR$ and long series of sunspot indices $SN$ ($ISN$ version 2.0) and sunspot groups $GN$ (Svalgaard and Schatten (2016) $GSN$ version). The Royal Greenwich observatory series after 1976 is extended by the Kislovodsk Mountain Astronomical Station data. When reconstructing $AR$ from $SN$, it is taken into account that the function $AR~=~f~(SN)$ has a non-linear systematic character and uncertainty associated with the heterogeneity of these indices. Therefore, in addition to modeling the most probable $AR$ values, predictive limits of reconstruction uncertainty are determined. In the interval 1610-1699 we carried the reconstruction out on the basis of the $GN$ series using the previously proposed decomposition in pseudo-phase space method (DPS). The resulting series NO21y is freely available online. We show that for this series the empirical Gnevyshev--Ohl rule and Waldmeier effect are fulfilled. Wavelet analysis reveals periodicities of 8.4-13.8 years for the main cycle (with a sharp decrease of the period before the global Maunder and Dalton minima) and a two-component Gleissberg cycle with typical periods of 50-60 years and 90-110 years.
\end{abstract}

\begin{keywords}
Sun: activity -- sunspots -- photosphere
\end{keywords}

%%%%%%%%%%%%%%%%%%%%%%%%%%%%%%%%%%%%%%%%%%%%%%%%%%

%%%%%%%%%%%%%%%%% BODY OF PAPER %%%%%%%%%%%%%%%%%%

\section{Introduction}\label{Intro}
Solar activity is represented by changes in the solar magnetic field at different spatial and temporal scales. The most famous and most tracked in time of its phenomenon is sunspots. They are mentioned in the chronicle sources of the last 2000 years, and attempts were made to create time series of solar activity on this basis \citep{Nagovitsyn2001, VaqueroGallegoGarcia2002}.

Telescopic observations of sunspots already cover more than 400 years; however, their systematic observations are fewer in time. Only since 1848 have Wolf numbers been satisfactorily monitored -- now they are called International Sunspot Numbers ($ISN$). Through the efforts of \citet{Cletteetal2014}, \citet{Cletteetal2015}, \citet{CletteLefevre2016}, we have a 320-year series of average annual values of this index (and monthly average values since 1749, \citet{SILSO}). It is usually called version 2.0, we will denote them as $SN$. Before the article by \citet{Cletteetal2014}, researchers used $ISN$ version 1.0 as a reference dataset when reconstructing various indices of solar activity \citep[see, for example,][]{Nagovitsyn2008}. A comparison of the two versions can be found in a number of works \citep[see, for example,][fig. 2]{Georgievaetal2017}.

\citet{HoytSchatten1998} proposed a series of group sunspot number ($GSN$) normalized to the Wolf number:
\begin{equation}
    GSN = \frac{12.08}{N} \sum k_{i}^{'} G_i,
	\label{eq1}
\end{equation}
where $G_i$ is the number of groups observed by the $i$\textsuperscript{th} observer, $k_{i}^{'}$ is the correction factor for the $i$\textsuperscript{th} observer, $N$ is the number of observations used to calculate the $GSN$. Currently, several versions of the group sunspot number have been proposed (we will denote them as $GN$): \citet{SvalgaardSchatten2016}, \citet{Usoskinetal2016}, \citet{CliverLing2016}, and \citet{Chatzistergosetal2017}. The longest (since 1610) is the \citet{SvalgaardSchatten2016} version, and we will use it in our work. It should be borne in mind that the data of the 17\textsuperscript{th} century are very scattered and obtained with the help of imperfect tools at that time. Plus, there are scattered light effects, observation gaps, and peculiar properties of the observers. All this forces us to treat these data with caution \citep{NagovitsynGeorgieva2017, KarachikPevtsovNagovitsyn2019}.

\begin{figure}
	\includegraphics[width=\columnwidth]{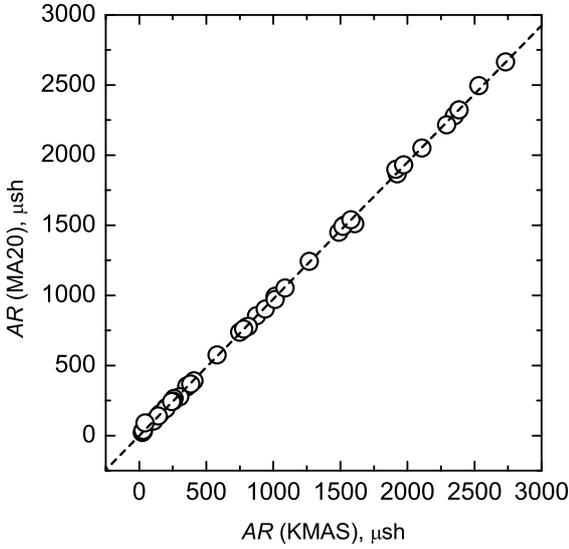}
    \caption{Comparison of the average annual total sunspot area index $AR$ by Kislovodsk mountain astronomical station (KMAS) and \citet{Mandaletal2020} (MA20) in 1977-2019.}
    \label{fig1}
\end{figure}

Another index of solar activity is the total sunspot area averaged over a selected period measured in $\mu sh$ (millionths of a (visible) solar hemisphere). We will denote this index averaged over a year as $AR$. This index is the most physically conditioned in comparison with the mentioned ones: it is associated with the absolute sunspot magnetic flux \citep{NagovitsynTlatovNagovitsyna2016}. Another advantage of the $AR$ index is that it weakly depends on the loss of small sunspots during observations \citep{NagovitsynGeorgieva2017}. The most famous and widely used is the Royal Greenwich observatory (RGO) $AR$ series, which existed from May 1874 to December 1976 \citep{BaumannSolanki2005}. Several attempts have been made to continue it. \citet{Balmacedaetal2009} used data from observatories of the former USSR in 1977-1985, USAF network in 1986-2009, and data from observatories in Rome, Yunnan, and Catania to fill the gaps. \citet{Foukal2014} showed that SOON (USAF network) data are not the best continuation of the Greenwich series. \citet{Mandaletal2020} compiled a continuation of the Greenwich series after 1976 (we denote this series as MA20) by combining data from the Kislovodsk Mountain Astronomical Station (KMAS), data from the Catalog of Solar Activity by R.S. Gnevysheva (Pulkovo Observatory) and data from the Debrecen Observatory \citep{BaranyiGyoriLudmany2016} as the closest to RGO sunspot area measurement systems. In previous works, we used only KMAS data for the continuation of the RGO series, relying on the fact that for the monthly mean values in the overlapping interval of RGO and KMAS in 1958-1976 the correlation coefficient is $r = 0.992$, and the correction factor is $k = 1.0094 \pm 0.0059$ \citep{NagovitsynTlatovNagovitsyna2016}. We add to this that for the same time interval, the average annual values that we will consider in this work correlate with $r = 0.995, k = 1.008 \pm 0.015$. Based on this, below we will continue the RGO series after 1976 with the original KMAS data, as in our previous works. Let us correlate the series of average annual values MA20 and KMAS in 1977-2019 -- see Figure \ref{fig1}.
The correlation coefficient is very high: $r = 0.9998$. The difference from 1 of the scaling coefficient $k = 0.9746 \pm 0.0018$ is most likely explained by the fact that \citet{Mandaletal2020} assumed $k = 0.98$ for the Kislovodsk data in the general series.

Now about the aims of this work. The main task is to compile the longest series of average annual total sunspot area and to find its possible range of values for different time intervals. We will denote this sought series as NO21y. For this, initially, using the long-term index $SN$ version 2.0 (1700-2020), we will reconstruct the index of average annual total sunspot areas $AR$, the observation period of which is shorter (from 1874 to 2020). Initially, it is clear that simple scaling will not help here: the $SN$ series describes the numerical aspect of sunspot formation (the number of spots plus the number of groups multiplied by 10), and $AR$ is associated with a physical quantity -- sunspot magnetic flux. Therefore, their relationship may be non-linear. Note, for example, that among large sunspots there are unipolar sunspots with Zurich class H, which can have a fairly large area. But in the $SN$ index H class sunspot will give only 11 units, i.e. the same as for small single spot. Therefore, it is clear that the dependence $AR~=~f~(SN)$ will also have a random (non-systematic) spread.

Nevertheless, on the one hand, we can get rid of non-linearity by statistical methods by compiling an $AR$ model based on $SN$, and on the other hand, we can estimate the (natural) uncertainty intervals of the resulting model associated with the heterogeneity of the indices. Thus, we need to build an $AR$ model with intervals of uncertainty of its possible values in the past.

Another task: \citet{delaRueStewartLoewy1870} measured the areas of the sunspots from sketches by Schwabe, Carrington, and their own photographic observations, in 1832-1868. Attempts have already been made to scale these observations into the RGO system \citep{VaqueroGallegoSanchezBajo2004, Carrascoetal2016}. Can we suggest a new approach to this procedure? What are the confidence intervals for such a reconstruction?

Note that \citet{Arlt2008} introduced Staudacher's observations over a long time interval from 1749 to 1796 into scientific circulation. It would be extremely interesting to `connect' these data to the $AR$ series. However, the examples of Staudacher’s sunspot sketches given by Arlt indicate that his observations are too schematic, and they can hardly be used by us.

11-year solar cyclic variability is characterized by several empirical rules relating the temporal evolution of the solar cycles with their strengths. These rules are derived from the analysis of observational data and can be used as constraints for solar dynamo models \citep{Charbonneau2020}. In this work, we consider two empirical rules: the Waldmeier effect \citep{Waldmeier1935, Waldmeier1939} which relates the length of the ascending phase of a solar cycle to its maximum amplitude, and the Gnevyshev--Ohl rule \citep{GnevyshevOhl1948}, which implies a strong connection between the integral intensities of an even and the subsequent odd cycles.

\section{Dependence of \textit{AR} from \textit{SN}}

\begin{figure}
	\includegraphics[width=\columnwidth]{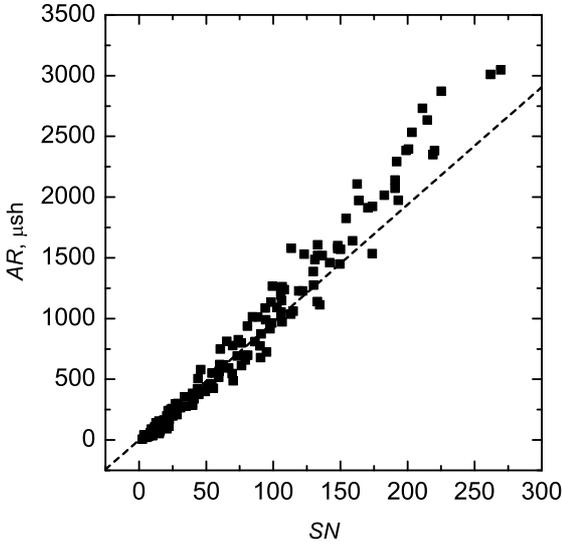}
    \caption{Comparison of the average annual total sunspot area index $AR$ by RGO (1875-1976) continued by KMAS (1977-2020) (see Section \ref{Intro} for explanation) and average annual international sunspot number $SN$ in 1875-2020. The dotted line is a linear dependence for $SN < 100$.}
    \label{fig2}
\end{figure}

Figure \ref{fig2} shows the dependence of the observed $AR$ on $SN$. The dotted line is a linear relationship for $SN < 100$. The non-linearity of the overall relationship is beyond doubt. Therefore, simple scaling of $SN$ to $AR$ is not possible.

To bring the values of different authors from \citet{delaRueStewartLoewy1870} to the Greenwich system, \citet{Carrascoetal2016} used the dependence of $AR$ on $SN$ in the form of a power-law:
\begin{equation}
   AR = a SN^b
   \label{eq2}
\end{equation}
This dependence (for monthly average values) was built according to the observational interval 1874-2009 and to the known $SN$ values. It was extrapolated to the interval 1832-1868, and then the correlation ratios given by \citet{delaRueStewartLoewy1870} $AR$ were constructed using the coefficients obtained from the approximation (\ref{eq2}). \citet{Carrascoetal2016} used monthly averages. In our work, we will use annual averages.

When determining the coefficients in equation (\ref{eq2}), we have two options. The first one is to linearize equation (\ref{eq2}), using the logarithm:
\vspace{-5pt}
\begin{equation}
   \ln AR = \ln a + b \ln SN,
   \label{eq3}
\end{equation}
find the coefficients using the least-squares method (LSM). In this case, we get
\vspace{-5pt}
\begin{equation}
    AR = (4.51 \pm 0.30) SN^{(1.177 \pm 0.016)}
    \label{eq4}
\end{equation}
The correlation coefficient $r$ between the obtained model $AR^*$ and the observables $AR^* = k AR$, the correction factor $k$, and the standard deviation $s$, in this case, will be
\begin{equation}
    r_1 = 0.988, \; k_1 = 0.9952 \pm 0.0086, \; s_1 = 119
    \label{eq5}
\end{equation}

The second option is the Levenberg--Marquardt algorithm \citep{Levenberg1944, Marquardt1963} (LMA), which allows working with non-linear equation \ref{eq2} directly (this method is implemented, in particular, in the \textsc{origin} package). Applying it, we get:
%\vspace{-5pt}
\begin{equation}
AR = (5.18 \pm 0.56)SN^{(1.149\pm0.021)},
\label{eq6}
\end{equation}
\vspace{-20pt}
\begin{equation}
r_2 = 0.988, \;  k_2 = 0.9994 \pm 0.0086, \;  s_2 = 118
\label{eq7}
\end{equation}

Another formula that can be used to linearize the dependence of $AR$ on $SN$ is the polynomial
\begin{equation}
AR = c SN + d SN^2
\label{eq8}
\end{equation}
\vspace{-12pt}

Since it is linear in terms of coefficients, we apply the LSM, we get:
\begin{equation}
AR = (9.08 \pm 0.29) SN + (0.0114 \pm 0.0017) SN^2
\label{eq9}
\end{equation}
\vspace{-20pt}
\begin{equation}
r_3 = 0.988, \;  k_3 = 1.0007 \pm 0.0088, \;  s_3 = 120
\label{eq10}
\end{equation}

All three approaches give similar results for the correlation between $AR$ and $SN$. In principle, we can choose any of them to linearize the model of the observed $AR$ dependence on the $SN$.

However, we will formulate another approach. \citet{NagovitsynPevtsov2016} have shown that the ratio of the number of large sunspot groups to small ones changes with an 11-year cycle: there are more large sunspot groups at the maximum. This means that the function  $AR~=~f~(SN)$  may depend on the phase (stage) of the cycle. Therefore, we divide the $SN$ dataset into stages according to the $n$ years from the maxima $M$ or minima $m$: $M + n$, where $n \in [-3, 5]$, and $m + n$, where $n \in [-5, 3]$. For each stage, we will seek its own dependence of $AR$ on $SN$. Let us choose equation (\ref{eq9}) as the simplest and most linear in terms of the fitting coefficients. Tables \ref{table1} and \ref{table2} show the values of the coefficients $c$ and $d$ of equation (\ref{eq8}) for different stages of the cycle, as well as the values of standard deviation $s$ and correlation coefficient $r$. We see that the non-linearity of the relationship between $AR$ and $SN$ manifests itself in different ways for different stages. It is most pronounced for stages $M~-~3, M~+~2, m~-~5, m~+~2$. For stages $M~-~2, M, M~+~4, m~-~3, m~-~1, m, m~+~1$, the function  $AR~=~f~(SN)$  is close to linear.
\begin{table}
\begin{tabular}{lcccl}
\hline
Stage $M + n$ & $c$             & $d \cdot 10^2$  & $s$ & $r$   \\ \hline
$M - 3$         & $6.7 \pm 1.0$   & $5.8 \pm 1.5$   & 42  & 0.991 \\
$M - 2$         & $7.2 \pm 1.9$   & $3.7 \pm 2.3$   & 113 & 0.926 \\
$M - 1$         & $7.4 \pm 1.5$   & $2.5 \pm 1.0$   & 161 & 0.950 \\ 
$M$             & $10.5 \pm 1.4$  & $0.40 \pm 0.69$ & 210 & 0.947 \\
$M + 1$         & $8.67 \pm 0.92$ & $1.10 \pm 0.47$ & 138 & 0.981 \\
$M + 2$         & $7.53 \pm 0.56$ & $2.29 \pm 0.31$ & 70  & 0.996 \\
$M + 3$         & $7.7 \pm 1.5$   & $2.6 \pm 1.1$   & 141 & 0.972 \\
$M + 4$         & $9.6 \pm 2.0$   & $0.4 \pm 2.4$   & 98  & 0.959 \\
$M + 5$         & $6.7 \pm 2.4$   & $6.0 \pm 4.8$   & 70  & 0.941 \\ \hline
\end{tabular}
\caption{Coefficients $c$ and $d$ of equation (\ref{eq8}) for different stages of the cycle divided by years $n$ of deviation from the maxima $M$. $r$ is the correlation coefficient, $s$ is the standard deviation.}
\label{table1}
\end{table}

\begin{table}
\begin{tabular}{lcccl}
\hline
Stage $m   + n$ & $c$             & $d \cdot 10^2$  & $s$ & $r$   \\ \hline
$m -5$          & $7.4 \pm 1.1$   & $2.33 \pm 0.63$ & 149 & 0.983 \\
$m - 4$       & $8.3 \pm 1.6$   & $2.1 \pm 1.2$   & 143 & 0.956 \\
$m-3$           & $7.5 \pm 1.5$   & $2.9 \pm 2.0$   & 88  & 0.945 \\
$m-2$           & $4.6 \pm 1.8$   & $9.8 \pm 3.6$   & 56  & 0.959 \\
$m-1$           & $8.7 \pm 1.8$   & $-2.1 \pm 6.6$  & 45  & 0.848 \\
$m$             & $5.4 \pm 2.5$   & $16 \pm 17$     & 26  & 0.878 \\
$m+1$           & $8.7 \pm 1.4$   & $0.7 \pm 3.2$   & 47  & 0.964 \\
$m+2$           & $9.19 \pm 0.70$ & $1.55 \pm 0.47$ & 88  & 0.992 \\
$m+3$           & $8.7 \pm 1.2$   & $1.22 \pm 0.61$ & 209 & 0.973 \\ \hline
\end{tabular}
\caption{Coefficients $c$ and $d$ of equation (\ref{eq8}) for different stages of the cycle divided by years $n$ of deviation from the minima $m$. $r$ is the correlation coefficient, $s$ is the standard deviation.}
\label{table2}
\end{table}

Now, knowing the stage of activity relative to the extremums, we can reconstruct $AR$ from $SN$ for the entire interval from 1700 to present. In the case of maxima and minima estimates intersection, we use the weighted average. We will call this reconstruction the W-version of $AR$. Correlating it with the observed values, we find:
\begin{equation}
r_4 = 0.990, \;  k_4 = 0.9985 \pm 0.0077, \;  s_4 = 106
\label{eq11}
\end{equation}

Comparing equation (\ref{eq11}) with (\ref{eq5}), (\ref{eq7}), (\ref{eq10}) we find that the W-version gets rid of the non-linearity in the function  $AR~=~f~(SN)$  slightly better. The residual standard deviation is 10 per cent less than in other cases (\ref{eq5}), (\ref{eq7}), (\ref{eq10}).

\section{\textit{AR} in 1875-2020 -- observational series in the RGO system}
\begin{figure*}
	\includegraphics[width=14cm]{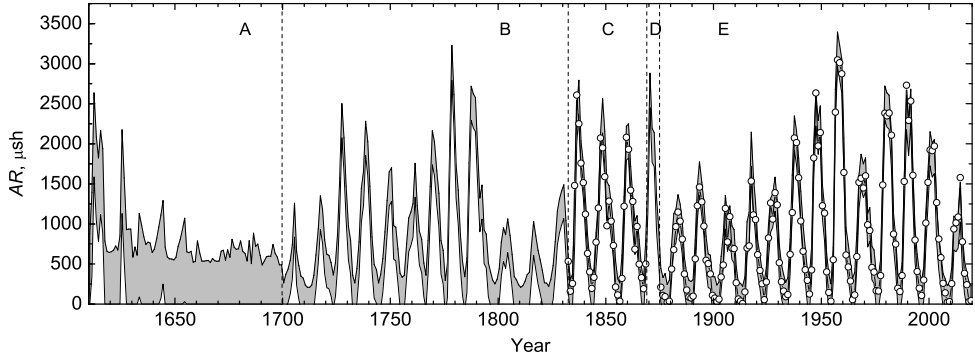}
    \caption{Gray areas of panel A correspond to 95 per cent prediction bands of the DPS-reconstruction (see Section 7). The gray areas for panels B-E correspond to 95 per cent prediction bands of the W-version. White circles are $AR$ observational series in the RGO system (de la Rue 1832-1868, RGO 1875-1976 and KMAS 1977-2020).}
    \label{fig3}
\end{figure*}
Figure \ref{fig3} is divided into panels A-E according to the type of sunspot area data available. The gray areas for panels B-E correspond to 95 per cent prediction bands \citep{Chatfield1993} of the W-version. We will refer to the final dataset obtained in this work as NO21y. Below we will describe the panels of Figure \ref{fig3} separately.

Panel E illustrates the alignment of the observed $AR$ values (white circles) with the W-version (gray areas on panels B-E). We see that the agreement is very good, which is confirmed by the above-found correlation coefficient $r_4~=~0.990$.

\section{\textit{AR} in 1832-1869 -- observational series by Schwabe, Carrington, and de la Rue}

\citet{delaRueStewartLoewy1870} provide the sunspot areas (averaged over half of a month) for the observations of Schwabe (1832-1853), Carrington (1854-1860), and their own at the Kew Observatory (1861-1868) (the latter are photographic). Having calculated the average annual values, let us compare them linearly with the W-version for the indicated intervals separately, denoting them with the $SH$, $CA$, and $DE$ indices according to the first two letters of the names of the observers. We get
\begin{equation}
r_{SH} = 0.955, \;  k_{SH} = 1.880 \pm 0.069, \;  s_{SH} = 228
\label{eq12}
\end{equation}
\vspace{-20pt}
\begin{equation}
r_{CA} = 0.996, \;  k_{CA} = 1.488 \pm 0.041, \;  s_{CA} = 86
\label{eq13}
\end{equation}
\vspace{-20pt}
\begin{equation}
r_{DE} = 0.970, \;  k_{DE} = 1.091 \pm 0.051, \;  s_{DE} = 111
\label{eq14}
\end{equation}

Panel C of Figure \ref{fig3} shows the data obtained using equations (\ref{eq12})-(\ref{eq14}) together with the W-version. For the NO21y estimate, we choose 95 per cent confidence bands.

\section{\textit{AR} in 1700-1831 and 1869-1874 -- W-version} 
For these time intervals, we can only rely on the W-version when evaluating $AR$. Since the obtained values are, in essence, a prediction of $AR$ values, the prediction bands of the W-version should be used for the estimates of the confidence bands in NO21y.

\section{\textit{AR} in 1610-1699 -- Svalgaard and Schatten (2016) series}

The $GN$ series of \citet{SvalgaardSchatten2016} contains data on the sunspot group number since the early 17\textsuperscript{th} century (we will denote it as $SG$). In the second half of this century, it is difficult to distinguish separate 11-year cycles, if they occur at all. Therefore, for the reconstruction of $SN$ earlier than 1700, we will go the other way.

In \citet{Nagovitsynetal2004}, a decomposition in terms of pseudo-phase space (DPS) reconstruction method was proposed. It follows from the approach of Takens (1981), who established, in particular, a relation of dynamical systems with autoregression models. Without going into the details of the method (it is described in detail in \citet{Nagovitsyn2008}, \citet{Nagovitsynetal2008}), we note that the shorter series (in our case $SN$) is decomposed into the components of the Takens pseudo-phase space formed by the shifts of the longer $GN$ series relative to itself by $\Delta, 2\Delta, 3\Delta, ..., (n-1) \Delta$. $\Delta$ is chosen near the first zero of the autocorrelation function. In our case, $\Delta = 3$ years. The number of components $n$ is usually taken equal to 7 \citep{Nagovitsyn2008}. As a result, we get a model of $SN$ behavior over a longer interval.

First, let us get rid of the non-linearity of the SG and $AR$ connection. Comparing these values, we get a new variable:
\begin{equation}
SG^* = (151.9 \pm 9.0)SG + (5.7\pm1.1)SG^2                               \label{eq15}
\end{equation}
Using it in the DPS method, we obtain model by the least-squares method for the interval 1610-1699 ($t$ is in years):
\begin{equation}
\begin{split}
AR(t) = 11 + 0.907 \cdot SG^*(t) + 0.033 \cdot SG^*(t + 3) - \\
 - 0.019 \cdot SG^*(t + 6) + 0.105 \cdot SG^*(t + 9) + \\
 + 0.060 \cdot SG^*(t + 12) - 0.041 \cdot SG^*(t + 15) - \\
 - 0.067 \cdot SG^*(t + 18)
\end{split}
\label{eq16}
\end{equation}
The correlation coefficient of model (\ref{eq16}) with the W-version $r = 0.932$, standard deviation $s = 267$. In Figure \ref{fig3}, panel A, the gray prediction bands are shown at the 95 per cent prediction interval. Note that the upper limit of the areas in the Maunder minimum almost corresponds to the average values in the Dalton minimum. This does not contradict the conclusions of \citet{ZolotovaPonyavin2015} on the possibility of non-anomalously low cycle amplitude in the Maunder minimum. But the cyclicity in the interval of 1660-1690 is barely expressed.

\section{Series of average annual total sunspot areas NO21y in 1610-2020}
\begin{figure*}
	\includegraphics[width=14cm]{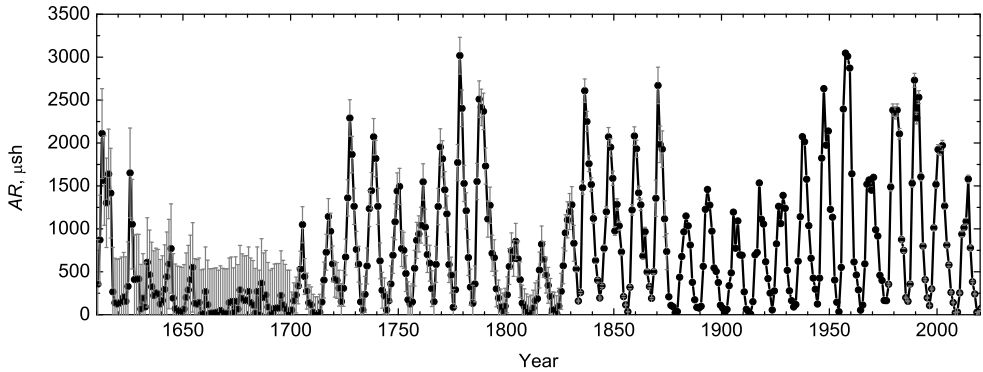}
    \caption{Constructed NO21y series of average annual total sunspot areas. This dataset is available at \url{http://www.gaoran.ru/database/csa/derived/AR_NO21y.txt}}
    \label{fig4}
\end{figure*}
Figure \ref{fig4} shows the NO21y series. In the interval 1610-1699, the values and prediction intervals (95 per cent prediction bands) were obtained by applying the DPS method (\ref{eq15})-(\ref{eq16}) to the \citet{SvalgaardSchatten2016} series. Time intervals 1700-1831 and 1869-1874 correspond to the values and prediction intervals (also 95 per cent prediction bands) of the W-version. The interval 1832-1868 and confidence intervals (95 per cent confidence bands) correspond to the processed observations of Schwabe, Carrington, and de la Rue. We assigned zero weights to the RGO series (1875-1976) as an observational and reference dataset. We took the original values of the KMAS series in 1977-2020, confidence intervals (95 per cent confidence bands) were obtained based on the comparison of RGO and KMAS in 1958-1976.

\section{NO21y series and empirical rules for solar cyclicity}
Among the regularities of solar cyclicity, as mentioned in the Introduction, there are two empirical rules: the Gnevyshev--Ohl rule \citep{GnevyshevOhl1948} and the Waldmeier effect \citep{Waldmeier1935, Waldmeier1939}. The latter also has a modified version \citep{NagovitsynKuleshova2012}. The rules were originally derived from a series of International Sunspot Numbers version 1.0. Recently \citet{UsoskinKovaltsovKiviaho2021} examined them for modern versions of $SN$ and $GN$.

Since we now have at our disposal a new series of $AR$ -- average annual total sunspot areas, which seem to be more physically-based than $SN$ and $GN$ -- it makes sense to consider the feasibility of these rules for it.

Since the 1610-1699 interval is problematic (see Figures \ref{fig3}, \ref{fig4}), we will consider these rules for the 320-year interval from 1700 to 2020.

\begin{figure*}
	\includegraphics[width=14cm]{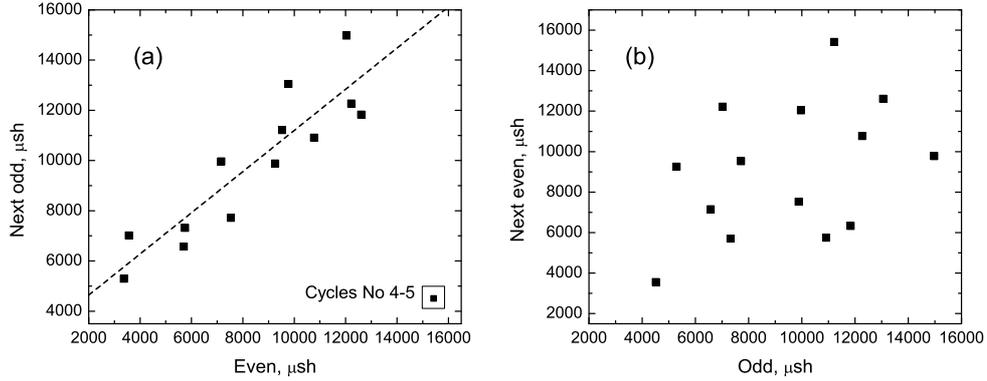}
    \caption{The Gnevyshev--Ohl rule for the NO21y series: the correlation of the integral intensity of an even cycle with a subsequent odd cycle (a); the correlation of the integral intensity of an odd cycle with a subsequent even cycle (b).}
    \label{fig5}
\end{figure*}

Figure \ref{fig5} shows the results obtained: on the left is the correlation of the integral intensity of an even cycle with a subsequent odd cycle, on the right -- conversely, an integral intensity of an odd cycle with a subsequent even cycle. In the first case, the correlation coefficient is $r = 0.90$, in the second $r = 0.39$. Following the classical work of \citet{GnevyshevOhl1948}, the pair of cycles of the Zurich numbering 4-5 is removed from statistics. Thus, the Gnevyshev--Ohl rule is also confirmed for the NO21y series: the cycles are physically linked precisely in the pairs of even and subsequent odd cycles.

\begin{figure*}
	\includegraphics[width=14cm]{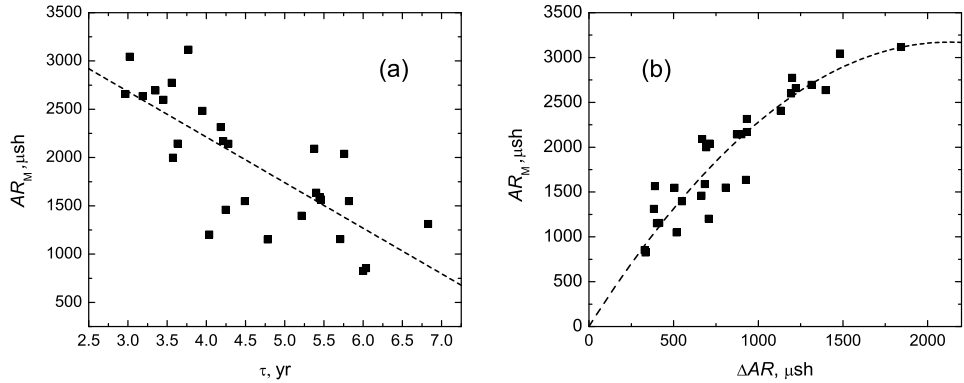}
    \caption{Waldmeier effect for the NO21y series. Classical Waldmeier effect: dependence of the maximum cycle amplitude on the length of the ascending phase of a solar cycle (a); modified Waldmeier effect: dependence of the maximum cycle amplitude on the highest rise rate on the ascending phase of a solar cycle (b).}
    \label{fig6}
\end{figure*}

Another empirical rule is the Waldmeier effect, which connects the length of the ascending phase of a cycle with the maximum cycle amplitude. To accurately determine the length of the ascending phase from the average annual data, we apply the following technique. Near the extremums of the average annual $AR$ value in year $i$, values in $i - 1$ and $i + 1$ years are selected, and an exact parabola passing through these points is constructed using the Lagrange method. The position of the extremum of this parabola and the corresponding value of $AR$ are considered the sought ones, and from them, the length of the ascending phase $t$ and the height of the cycle at the maximum $AR_M$ are calculated. Figure \ref{fig6}a shows the resulting dependence. The linear correlation coefficient is $r = -0.77$. The correlation is quite low, but it is exists, and this is typical for the Waldmeier effect, considered for other indices \citep{UsoskinKovaltsovKiviaho2021}.

\citet{NagovitsynKuleshova2012} considered a modification of the Waldmeier effect, namely, the dependence of the maximum cycle amplitude on the highest rise rate on the ascending phase of a solar cycle, in this case, $\Delta AR$. Figure \ref{fig6}b shows the resulting dependence, which, as in the article by \citet{NagovitsynKuleshova2012}, was approximated by a quadratic polynomial (with the obvious condition $\Delta AR = 0$ at $AR_M = 0$):
\begin{equation}
\begin{split}
AR_M = (2.98 \pm 0.16) \Delta AR - (7.0 \pm 1.3) \cdot 10^{-4}(\Delta AR)^2, \\
r = 0.929
\end{split}
\label{eq17}
\end{equation}
Thus, the modified Waldmeier effect for the NO21y series is also fulfilled: the correlation is quite high.

\section{Cycles of solar activity according to NO21y data}
\begin{figure}
	\includegraphics[width=9cm]{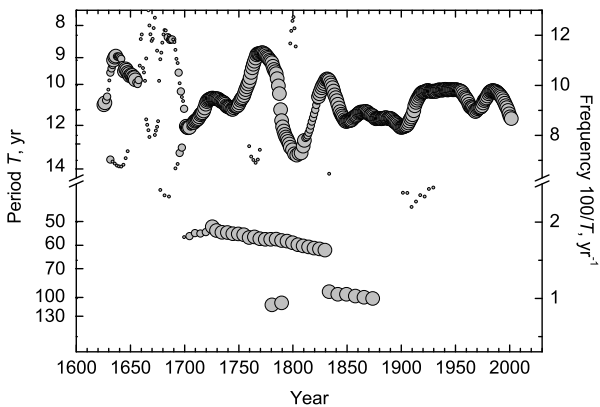}
    \caption{Wavelet transform skeleton (the positions of the local maxima of the wavelet spectrum) for NO21y series. The largest circles correspond to reliability $>~99$ per cent, medium -- $95-99$ per cent, the smallest $<95$ per cent.}
    \label{fig7}
\end{figure}
Let us consider the changes in the cyclicity manifested in the NO21y series. For this, we use the approach of \citet{GelfreikhNagovitsynNagovitsyna2006} based on the Morlet wavelet of the 6\textsuperscript{th} order \citep{Astafeva1996}. Figure \ref{fig7} shows the skeleton -- the positions of the local maxima of the wavelet spectrum. The largest circles correspond to reliability $>99$ per cent, medium -- $95-99$ per cent, the smallest $<95$ per cent. The main (high-frequency) component, corresponding to the 11-year cycle, varies within 8.4-13.8 years. The shortest periods are observed before the global Maunder and Dalton minima. At the Maunder minimum, the cyclicity is weakly expressed, but a local decrease in the period in its middle is observed. The longest period -- 13.8 years -- is in the Dalton minimum. At present, the behavior of a period similar to the one preceding the global minimum is not observed. Apparently, a decrease in the amplitude of the last cycles will not lead to minima of the Maunder or Dalton type \citep[see also][]{Hayakawaetal2020}.

Figure \ref{fig7} looks interesting in the low-frequency part of the wavelet spectrum. Instead of the Gleissberg cycle of 80-90 years, we see a bimodal picture. One of these long cycles lasts 50-60 years, the other 90-110 years. Note that such a bimodal structure has already been found in other data: Figure 5 in \citet{Nagovitsyn2001} and Figure 5 in \citet{Nagovitsynetal2004}.

\section{Summary and conclusions}
The average annual total sunspot area index $AR$ is physically conditioned in contrast to the `statistical' indices of the relative sunspot number $SN$ and groups $GN$ since it is associated with the sunspot magnetic flux. Unfortunately, the observed $AR$ dataset is much shorter than $SN$ and $GN$. The aim of this work was a long reconstruction of the $AR$ series using $SN$ and $GN$.

We confirmed that the KMAS data can be used as a continuation of the RGO sunspot areas $AR$ dataset after 1976.

It is a priori clear that the dependence  $AR~=~f~(SN)$  contains systematic and fluctuation components. The systematic component is non-linear \citep{Carrascoetal2016}. The fluctuation component is associated with the heterogeneity of the $AR$ and $SN$ indices: if $AR$ reflects the magnetic flux of the sunspots, then $SN$ reflects the structure and number of sunspots in the active region. Therefore, it is necessary to correct the non-linearity, and from the scatter of points between the model and real observations, estimate the range of possible values of the $AR$ series reconstructed from $SN$ for times when there were no $AR$ observations.

We show that to correct the non-linearity of the dependence  $AR~=~f~(SN)$, different types of functions can be used, but the best is the dependence that takes into account the stage of the cycle relative to the extremums. On this basis, we build the so-called W-version of the $AR$, based on the $SN$ series version 2.0.

We converted the data of Schwabe, Carrington, and de la Rue to the RGO dataset system. We found the most probable values and prediction bands for the average annual total sunspot areas starting from 1610. The data in the 17\textsuperscript{th} century are estimated since at that time there were not enough regular observations of solar activity.

The obtained NO21y series demonstrate the feasibility of the empirical Gnevyshev--Ohl rule and Waldmeier effect starting from 1700.

The wavelet transform skeleton describing the changes in the local periods of the cyclicity for the NO21y series indicates a decrease in the periods before global minima, as well as the bimodality of the Gleissberg cycle.

We see further development of this work in the compilation of a long series of monthly average total sunspot areas.

\section*{Acknowledgements}
This work was supported by the Program of Large Projects of the Ministry of Science and Higher Education of the Russian Federation (grant No 075-15-2020-780).

\section*{Data Availability}
The resulting NO21y series is available at \url{http://www.gaoran.ru/database/csa/derived/AR_NO21y.txt}.

%%%%%%%%%%%%%%%%%%%%%%%%%%%%%%%%%%%%%%%%%%%%%%%%%%



%\bibliographystyle{mnras}


% Don't variation these lines
\bsp	% typesetting comment
\label{lastpage}
\end{document}